\documentstyle{mn}

\newif\ifAMStwofonts
\AMStwofontstrue

\input psfig

\newcommand{\cmcu}{\,\hbox{cm}$^3$}
\newcommand{\pcmcu}{\,\hbox{cm}$^{-3}$}
\newcommand{\kms}{\,\hbox{\hbox{km}\,\hbox{s}$^{-1}$}}
\newcommand{\wsqm}{\,\hbox{\hbox{W}\,\hbox{m}$^{-2}$}}
\newcommand{\wsqma}{\,\hbox{\hbox{W}\,\hbox{m}$^{-2}$\,\hbox{arcsec}$^{-2}$}}
\newcommand{\msun}{\,\hbox{M$_{\odot}$}}
\newcommand{\lsun}{\,\hbox{L$_{\odot}$}}
\newcommand{\av}{\hbox{A$_{\rm V}$}}
\renewcommand{\deg} {\hbox{$^\circ$}}
\newcommand{\um}{\,\hbox{$\mu$m}}
\newcommand{\ha}{H$\alpha$}
\newcommand{\brg}{Br$\gamma$}

\def\s1{1-0\,S(1)}
\def\S1{1-0\,S(1)\,$\lambda$\,2.12\,$\mu$m}

\newcommand{\hiiregions}{\hbox{H\,{\sc ii}~regions}}
\def\h2{H$_2$}
\def\spose#1{\hbox to 0pt{#1\hss}}
\def\simlt{\mathrel{\spose{\lower 3pt\hbox{$\mathchar"218$}}
     \raise 2.0pt\hbox{$\mathchar"13C$}}}
\def\simgt{\mathrel{\spose{\lower 3pt\hbox{$\mathchar"218$}}
     \raise 2.0pt\hbox{$\mathchar"13E$}}}
\def\flux18{\nobreak{$\times10^{-18}$\,W}\,m$^{-2}$}
\def\flux17{\nobreak{$\times10^{-17}$\,W}\,m$^{-2}$}
\def\plotfiddle#1#2#3#4#5#6#7{\centering \leavevmode
    \vbox to#2{\rule{0pt}{#2}}
    \includegraphics{#1}}

\ifoldfss
  \ifCUPmtlplainloaded \else
    \NewTextAlphabet{textbfit} {cmbxti10} {}
    \NewTextAlphabet{textbfss} {cmssbx10} {}
    \NewMathAlphabet{mathbfit} {cmbxti10} {} 
    \NewMathAlphabet{mathbfss} {cmssbx10} {} 
  \fi
  \ifAMStwofonts
    \ifCUPmtlplainloaded \else
      \NewSymbolFont{upmath} {eurm10}
      \NewSymbolFont{AMSa} {msam10}
      \NewMathSymbol{\upi}     {0}{upmath}{19}
      \NewMathSymbol{\umu}     {0}{upmath}{16}
      \NewMathSymbol{\upartial}{0}{upmath}{40}
      \NewMathSymbol{\leqslant}{3}{AMSa}{36}
      \NewMathSymbol{\geqslant}{3}{AMSa}{3E}

      \let\leq=\leqslant \let\le=\leqslant
       \let\ge=\geqslant
    \fi
  \fi
\fi 

\ifnfssone
  \newmathalphabet{\mathit}
  \addtoversion{normal}{\mathit}{cmr}{m}{it}
  \addtoversion{bold}{\mathit}{cmr}{bx}{it}
  \newmathalphabet{\mathbfit} 
  \addtoversion{normal}{\mathbfit}{cmr}{bx}{it}
  \addtoversion{bold}{\mathbfit}{cmr}{bx}{it}
  \newmathalphabet{\mathbfss} 
  \addtoversion{normal}{\mathbfss}{cmss}{bx}{n}
  \addtoversion{bold}{\mathbfss}{cmss}{bx}{n}
  \ifAMStwofonts
    \ifCUPmtlplainloaded \else
      \UseAMStwoboldmath
      \makeatletter
      \new@mathgroup\upmath@group
      \define@mathgroup\mv@normal\upmath@group{eur}{m}{n}
      \define@mathgroup\mv@bold\upmath@group{eur}{b}{n}
      \edef\UPM{\hexnumber\upmath@group}
      \new@mathgroup\amsa@group
      \define@mathgroup\mv@normal\amsa@group{msa}{m}{n}
      \define@mathgroup\mv@bold\amsa@group{msa}{m}{n}
      \edef\AMSa{\hexnumber\amsa@group}
      \makeatother
      \mathchardef\upi="0\UPM19
      \mathchardef\umu="0\UPM16
      \mathchardef\upartial="0\UPM40
      \mathchardef\leqslant="3\AMSa36
      \mathchardef\geqslant="3\AMSa3E

      \let\leq=\leqslant \let\le=\leqslant
       \let\ge=\geqslant
    \fi
  \fi
\fi 

\ifnfsstwo
  \DeclareMathAlphabet{\mathbfit}{OT1}{cmr}{bx}{it}
  \SetMathAlphabet\mathbfit{bold}{OT1}{cmr}{bx}{it}
  \DeclareMathAlphabet{\mathbfss}{OT1}{cmss}{bx}{n}
  \SetMathAlphabet\mathbfss{bold}{OT1}{cmss}{bx}{n}
  \ifAMStwofonts
    \ifCUPmtlplainloaded \else
      \DeclareSymbolFont{UPM}{U}{eur}{m}{n}
      \SetSymbolFont{UPM}{bold}{U}{eur}{b}{n}
      \DeclareSymbolFont{AMSa}{U}{msa}{m}{n}
      \DeclareMathSymbol{\upi}{0}{UPM}{"19}
      \DeclareMathSymbol{\umu}{0}{UPM}{"16}
      \DeclareMathSymbol{\upartial}{0}{UPM}{"40}
      \DeclareMathSymbol{\leqslant}{3}{AMSa}{"36}
      \DeclareMathSymbol{\geqslant}{3}{AMSa}{"3E}

      \let\leq=\leqslant \let\le=\leqslant
       \let\ge=\geqslant
    \fi
  \fi
\fi 

\ifCUPmtlplainloaded \else
  \ifAMStwofonts \else 
    \def\upi{\pi}
    \def\umu{\mu}
    \def\upartial{\partial}
  \fi
\fi

\title[The Circumnuclear Environment of NGC\,1068]
{Star Formation in the Circumnuclear Environment of NGC\,1068}
\author[R. I. Davies, H. Sugai, and M. J. Ward]
         {Richard I. Davies,$^1$\thanks{Current address:
Max-Planck-Institut f\"ur extraterrestrische Physik, Postfach 1603,
85740 Garching, Germany \newline email: davies@mpe.mpg.de}
Hajime Sugai$^2$ and Martin J. Ward$^3$\\
         $^1$ Astrophysics, Oxford University, Nuclear Physics
         Building, Keble Road, Oxford, OX1 3RH, UK \\        
         $^2$ Department of Astronomy, Kyoto University, Kyoto 606-01,
         Japan \\
         $^3$ Department of Physics and Astronomy, Leicester
         University, Leicester, LE1 7RH, UK}

\date{Accepted 1988 December 15.
      Received 1988 December 14;
      in original form 1988 October 11}

\pagerange{\pageref{firstpage}--\pageref{lastpage}}
\pubyear{1994}

\begin{document}

\maketitle

\label{firstpage}

\begin{abstract}
We present near-infrared emission line images of the circumnuclear
ring in NGC\,1068.
We have measured the Br$\gamma$ fluxes in a number of star forming
complexes and derived extinctions for each of these by comparison with
H$\alpha$.
We investigate the star forming histories of these regions and find
that a short burst of star formation occured co-evally throughout the
ring within the last 30--40\,Myr, and perhaps as recently as 4--7\,Myr
ago.
The 1-0\,S(1) flux and S(1)/Br$\gamma$ ratios indicate that as well as
fluorescence, shock excited H$_2$ emission contributes to the total flux.
There is excess H$_2$ flux to the North-West where the ionisation cone
crosses the ring, and we have shown it is possible that the non-stellar
continuum from the Seyfert nucleus which produces the high excitation
lines could also be causing fluorescence at the edges of molecular
clouds in the ring.
The nuclear 1-0\,S(1) is more extended than previously realised but
only along the bar's major axis, and
we consider mechanisms for its excitation.
\end{abstract}


\section{Introduction}
\label{n1068:sec:intr}

NGC\,1068 is considered an archetypal Seyfert\,2 galaxy,
and being nearby and bright (14.4\,Mpc and $L_{\rm IR} = 2 \times
10^{11}$\lsun, Ringberg Standard) has been observed extensively.
In recent years much effort has been devoted to studying the nucleus at
ever higher spatial resolutions, exemplified by a direct 8.4\,GHz image of
the inner torus at a resolution of 2\,milli-arcsec (Gallimore, Baum \&
O'Dea, 1997), and 
adaptive optics imaging (Thatte et al. 1997) which showed that
94~per cent of the K-band light in the central 1\,arcsec originates
from a $\leq 30$\,milli-arcsec source, the remainder being due to a nuclear
star cluster.

But the circumnuclear ring at a radius of 15--16\,arcsec should not be
overlooked as here the \hiiregions, equivalent
to a string of M\,82-type galaxies, rival the nucleus in terms of
bolometric luminosity.
Telesco \& Decher (1988) explained the ring in terms of inner Lindblad
resonances associated with a barred potential, and Scoville et
al. (1988) showed that there was also a bar extending from this ring
closer in towards the nucleus.
Such a hierarchical structure is one of the most efficient ways in
which to transport gas over the 3--4 orders of magnitude in scale from
the disk, not only into the circumnuclear regions, but into the nucleus
itself.
Understanding what processes are operating in the ring, and how and why
they do so is a necessary step towards a global picture of NGC\,1068.
Atherton, Reay \& Taylor (1985) presented a velocity resolved H$\alpha$
image of the ring and circumnuclear environment, finding that there
were a large number of clouds with highly turbulent motions,
and possibly a large scale expansion of the disc.
This expansion was also seen by Planesas, Scoville \& Myers (1991) who
made high resolution CO observations.
The CO observations revealed a large number of cloud complexes in the
ring and nucleus, which appear to be much warmer ($\sim$50\,K) and
more massive (up to several $\times 10^8$\msun) than molecular clouds
in the Galaxy.

The inner disk has one of the highest blue surface brightnesses known
(Keel \& Weedman, 1978) and is the origin for about 1/3 as much \ha\ 
again as the ring.
Bland-Hawthorn, Sokolowski \& Cecil (1991) argue that this diffuse
ionised medium, characterised by a high [N\,{\sc ii}]/H$\alpha$ ratio,
is excited by scattered nuclear radiation, whereas 
the high excitation lines (Evans \& Dopita 1986; Bergeron 1989) arise
in gas which has a direct line of sight to the Seyfert nucleus.
The morphology of the high excitation gas has been studied in detail by
Pogge (1988) and 
Unger et al. (1992), who traced it out beyond the ring at PA 11--51\deg.
An ionisation cone such as this would naturally occur as a direct
result of collimation by a molecular torus, but simple interpretation of the
observations in terms of this model is not possible.
Unger et al. demonstrate that a proper analysis requires careful study of the
complex geometries of the radio jet, [O\,{\sc iii}] cone, UV field, etc.

Because of its brightness, proximity and favourable orientation,
NGC\,1068 provides an excellent laboratory for studying the Seyfert
nucleus, inner disk and circumnuclear ring.
Such work in many different wavebands (see the collage in Bland-Hawthorn
et al. 1997) is already providing insights into how these phenomena are
inter-linked, a topic which is receiving considerable attention at
present.
One aspect which has been previously omitted is the circumnuclear
near-infrared line emission and in this paper we present the first such
1-0\,S(1) and \brg\ images, which have been made possible because of the
wide field available with the Fabry-Perot and IRCAM3 on UKIRT.
We study individual star forming complexes, and by
comparison of the recombination line strengths determine the
extinctions to each region.
We look at the timescales for the star formation, 
determining whether it is co-eval throughout the ring and if there
have been previous episodes of activity in
Section~\ref{n1068:sec:time}.
In Section~\ref{n1068:sec:sf} we consider triggering mechanisms with
reference the effect of the ionisation cone and radio jet, and in
Section~\ref{n1068:sec:nucleus} we discuss the origin of the 1-0\,S(1)
emission in the ring.
Our signal-to-noise is sufficient to have revealed H$_2$ emission near
the nucleus which is more extended than was previously realised, and in
Section~\ref{n1068:sec:inner} we consider what may be the cause of
this.
We summarise our conclusions in Section~\ref{n1068:sec:conc}.

\section{Observations and Data Reduction}
\label{n1068:sec:obs}

Images of the 2\um\ continuum, and H$_2$\,1-0\,S(1) and Br$\gamma$
emission lines were obtained with the UKIRT 3.8-m telescope on the nights
of 19 and 20 December 1996, using the tip-tilt facility.
A Fabry-Perot etalon, with spectral resolution of 325\kms, was used in
conjunction with IRCAM3 at the Cassegrain focus.
IRCAM3 is a near-infrared camera housing a $256 \times 256$ InSb array
with a scale of 0.3\,arcsec per pixel, giving an unvignetted field of
view of more than 60\,arcsec.
Cooled narrow band filters (FWHM 2.4~per cent) were used for order
sorting, and wavelength calibration was achieved by peaking up on the
2.1171\um\ line from a Krypton lamp.

The observations were carried out as follows.
For each line, 2 or 3 minute observations of the object were made at the
on-line wavelength and also of the continuum at slightly shorter and
longer wavelengths (shifted with respect to the line centre by
$\pm1200$\kms\ for 1-0\,S(1), and by $-1200,+1080$\kms\ for Br$\gamma$).
Sky frames were also made at each wavelength.
This was repeated until sufficient on-line integration had been
achieved.

IRAF was used for all the image processing and analysis as described
below.
A flatfield was made from the bias-subtracted 3-minute sky-frames.
This was then divided into the object frames after each corresponding
sky frame had been subtracted.
Frames were registered on the bright peak of emission from the Seyfert
nucleus.
Each frame was then flux calibrated.
Non-photometric conditions on the first night and saturation of the
star used for Br$\gamma$, meant that only the
1-0\,S(1) line on the second night could be calibrated directly from a
standard star (HR\,1091, an A1 star with no spectral features at the
wavelengths used).
All the other frames were self-calibrated; that is, the continuum
(sampled only in regions with no line emission) was scaled to match the
calibrated frames.
This was feasible because the continuum emission in NGC\,1068
is bright over a sufficiently extended area.

Multiple ghost images were a severe problem with this object.
However, as the location of the ghosts was different for the two nights
on which the observations were made, these could be masked out.
This was carried out while co-adding the individual calibrated frames.
Finally, the continuum frame was subtracted from the on-line frames to leave
ghost-free line-only images.
The total on-line integration times were 42\,minutes for Br$\gamma$ and
30\,minutes for 1-0\,S(1).

The resolution (including the effects of seeing and registering, as
well as the enhancement afforded by the tip-tilt) was measured from
the standard star profiles to be 1.0\,arcsec.
Unfortunately, as the emission is extremely faint, the images in
Figure~\ref{fig:n1068:morph} have had to be smoothed with a median
filter to enhance the signal-to-noise, 
reducing the resolution to either 1.4 or 1.7~arcsec as shown.

\subsection{Calibration \& the Velocity Field}
\label{n1068:ss:vel}

One important point to consider is whether the calibration is significantly
affected by the velocity distribution of the line emitting gas itself,
since Atherton et al. (1985) showed that the line-of-sight velocities
of H$\alpha$ in the circumnuclear ring varied over 1020--1260\kms.
For our observations we centred the FP at 1136\kms, corresponding
to a heliocentric velocity of 1114\kms, so that  
the 325\kms\ resolution of the FP covered the range 950--1280\kms.
However, there is a shift in transmitted wavelength (towards shorter
wavelengths) which increases to larger radius, reaching
$\delta\lambda \equiv 160$\kms\ at $r = 30$\,arcsec.
Since $\delta\lambda \propto r^2$, we estimate that in the ring
$\delta\lambda \equiv 40$\kms.
We have taken this into account when correcting fluxes for the
variations in velocity.
The correction required has been calculated as $1/A(v)$ where $A(v)$ is the
value of the Airy function at a velocity $v$ offset from peak transmission.
The velocity dispersion of the lines has not been included since it
would require detailed knowledge of the line profiles across the entire
field, and  
would only entail an additional correction varying from almost
zero to a maximum of 10~per cent in the South-West where $\sigma>135$\kms.

The approximate velocities of the Br$\gamma$
components marked in Fig.~\ref{fig:n1068:morph} are given in
Table~\ref{tab:n1068:calib}, together with the correction factor.
Throughout the rest of the paper, the observed fluxes given have been
velocity corrected.
The whole image has not been corrected as, instead of simplifying its
interpretation, it would serve only to introduce a variable detection
limit across the field.

\begin{table}
\flushleft
\caption{Flux Correction Factors \label{tab:n1068:calib}}
\begin{tabular}{lrrrr}

\\
Region & \multicolumn{2}{l}{Position (arcsec} & Velocity & Correction
\\
& \multicolumn{2}{l}{from nucleus)} & \kms & factor \\

\\

A & 11.4\,E & 10.2\,N & 1050 & 1.01 \\
B & 9.1\,E  & 12.9\,N & 1090 & 1.02 \\
C & 6.0\,E  & 13.6\,N & 1120 & 1.11 \\
D & 3.7\,E  & 15.3\,N & 1130 & 1.16 \\
E & 5.2\,W  & 15.3\,N & 1220 & 1.90 \\
F & 13.2\,W & 7.2\,S  & 1270 & 2.57 \\
G & 10.6\,W & 10.7\,S & 1260 & 2.41 \\
H & 7.4\,W  & 13.1\,S & 1200 & 1.68 \\
I & 4.5\,W  & 14.3\,S & 1180 & 1.49 \\
J & 1.7\,W  & 13.3\,S & 1160 & 1.33 \\
K & 1.7\,E  & 13.0\,S & 1140 & 1.21 \\
L & 6.8\,E  & 11.6\,S & 1060 & 1.00 \\
M & 19.2\,E & 3.4\,N  & 1010 & 1.12 \\
N & 20.5\,E & 9.0\,N  & 1030 & 1.05 \\
O & 20.8\,E & 12.1\,N & 1030 & 1.05 \\

\\

\end{tabular}

\end{table}

\section{Near-Infrared Line Morphology}
\label{n1068:sec:morph}

\begin{figure*}
\plotfiddle{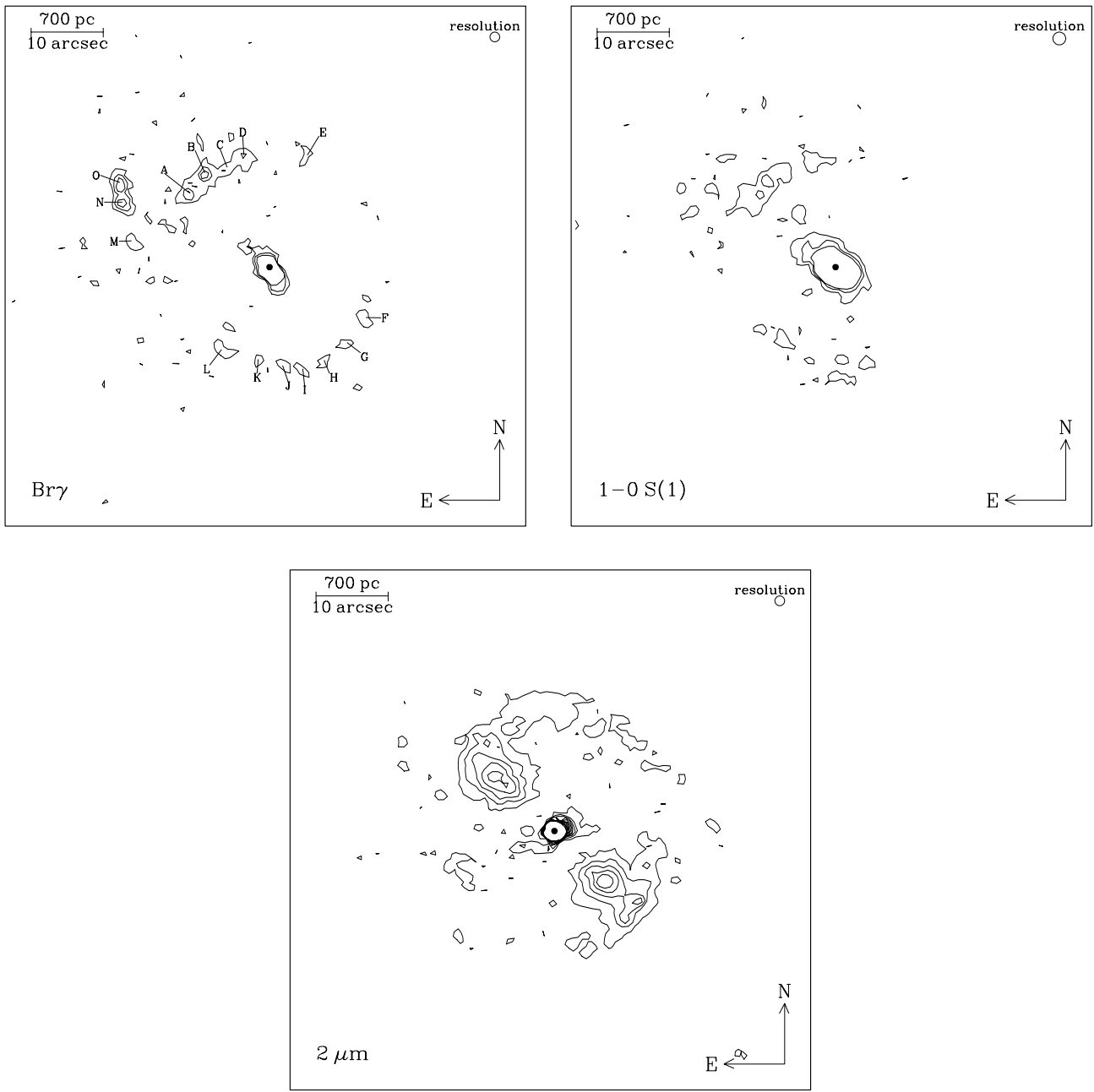}{17cm}{0}{98}{98}{-290}{-280}
\caption{Contour plots of a $72 \times 72$\,arcsec$^2$ field at the centre
of NGC\,1068.
The calibration shown in the line images has not been corrected for the
velocity field.
Upper Left -- Br$\gamma$ line emission, with contours 
at 2.9, 5.7 \& 8.6 $\times 10^{-19}$\wsqma;
the lowest contour is 2.5$\sigma$.
Upper right -- 1-0\,S(1) line emission, with contours at the same
intervals;
the lowest contour is 3$\sigma$ due to the greater smoothing used.
Lower -- 2\um\ continuum after subtraction of best-fit ellipses, with
contours from 0.1--2.3\,mJy\,arcsec$^{-2}$ at intervals of
0.2\,mJy\,arcsec$^{-2}$;
the lowest contour is 2.5$\sigma$.
The central (peak) region has not been contoured and instead the
location of the nucleus is marked by a large dot.}
\label{fig:n1068:morph}
\end{figure*}

Images of the Br$\gamma$ and 1-0\,S(1) line emission, and 2\um\ 
continuum are presented in Figure~\ref{fig:n1068:morph}.
The central regions have not been contoured;
instead the location of the nucleus, as defined by the centroid of the
continuum emission, has been marked with a large dot.

\subsection{2\um\ continuum}

Calibration of the continuum in several annuli compares favourably to
that in Scoville et al. (1988).
The emission is dominated by elliptical isophotes, and in order to
better recover the fainter underlying structure, these have
been modelled and subtracted using tasks in the ISOPHOTE package
within IRAF.
In order to preserve this structure, its inclusion
in the fitting process was avoided by using a clipping algorithm which
removed the 40~per cent of pixels with the highest counts before
fitting each isophote.
The bar is very clear lying along position angle 46\deg, as has been measured
by other authors (eg Scoville et al. 1988).
At the regions of peak intensity in the residual image, the bar has a
surface brightness approximately 75~per cent of the underlying
elliptical isophotes.
Closer to the nucleus, the clipping algorithm may not have been
efficient in excluding pixels in the bar due to the small size of the
ellipses and width of the bar.
We therefore do not consider the apparent lack of structure close to
the nucleus to be significant;
indeed the H$_2$ emission suggests that the bar does continue much
further in (see also Section~\ref{n1068:sec:inner}).
All the remaining structure in the residual image,
particularly at the southern end of the bar and the spiral arm west of
the northern end of the bar, also appears in the original continuum image.

It is interesting that the strongest continuum in the ring follows the
10\um\ morphology rather than that indicated by \ha\ or \brg\ knots,
suggesting that the emission mechanisms may be associated. 

\subsection{Molecular Hydrogen \s1 }

\begin{figure}
\psfig{file=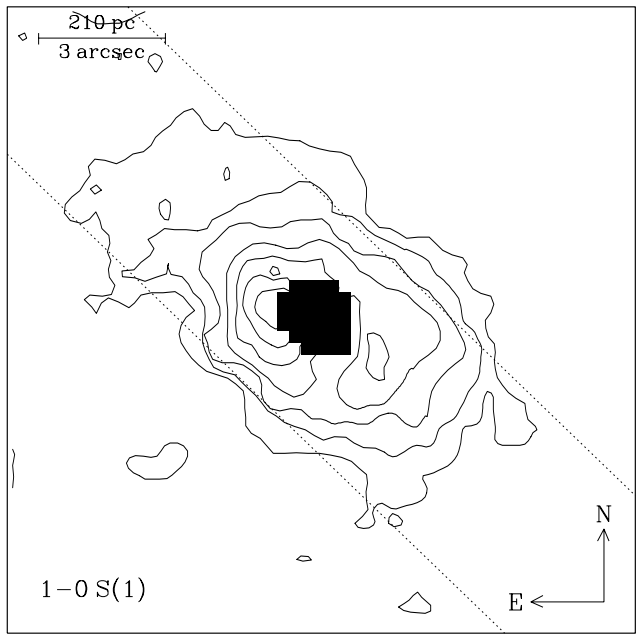,width=8.4cm,rheight=8.4cm}
\caption{Contour plot of 1-0\,S(1) in NGC\,1068.
Contours are from $3.6 \times 10^{-19}$\wsqma\ ($3.8\sigma$), with
levels increasing by a factor of 2, to $229 \times 10^{-19}$\wsqma.
These have not been corrected for the 250\kms\ line width.
The lowest levels have been smoothed to a resolution of 1.7\,arcsec,
while the highest levels are at the full resolution of 1.0\,arcsec.
The parallel dotted lines are the same as those shown in
Fig.~3 and delineate the edges of the bar as derived
from the 3rd contour in the continuum image.
Those pixels near the nucleus which are above some threshold in the
on-line image have been blacked out, as these are most affected by
saturation.}
\label{fig:n1068:inner}
\end{figure}

The 1-0\,S(1) emission is relatively strong at the end of the bar to the
North-East, as well as projecting from the nucleus along the first few
arcsecs of the bar.
The morphology in the central few arcsecs is consistent with that observed
by Rotaciuc et al. (1991) and Blietz et al. (1994).
The nuclear region shows a strong peak almost due East of the nucleus
and a weaker one to the South-West (Fig.~\ref{fig:n1068:inner}), but
these authors did not observe the more extended nuclear emission  as
their data went less deep.
As a rough consistency check of the total flux in a $6 \times 6$\,arcsec$^2$
aperture, we first interpolated across the pixels badly affected by the
saturation, and then applied a 20~per cent correction for the 200\kms\
line width.
The resulting flux of $13.8\pm0.1 \times 10^{-17}$\wsqm\ agrees well
with that measured by Oliva \& Moorwood (1990) in a similar aperture,
and also Blitz et al. (1994).

Reay, Atherton, and Walton (1986) previously detected \s1\ at two
points in the ring using a spectrometer with a 5.4\,arcsec aperture.
At the North-East end of the bar they measured an intensity almost an
order of magnitude greater than ourselves.
We believe that large error bars (points on the spectrum are 2--3$\sigma$
above their continuum level) combined with an underestimate of the
continuum level, have led to an overestimate of the line flux.
Their nuclear flux, detected at a much higher signal-to-noise,
is consistent with our measurement.

\subsection{Hydrogen Recombination \brg}

The Br$\gamma$ clearly delineates sections of the circumnuclear
ring, showing distinct similarities to that seen in the H$\alpha$ line
map (Bland-Hawthorn, Sokolowski \& Cecil, 1991).
Although the strongest emission appears to be to the North-East, this
is partly due to the effects of the velocity field as described in
Section~\ref{n1068:ss:vel}.
Table~\ref{tab:n1068:prop} shows that the emission to the South-West is
as strong as that in the North-East section of the ring.
The relative apparent weakness of the emission to the West is also due to
velocity effects, with the result that any
Br$\gamma$ emission is below our detection threshold.
In the nucleus there is evidence for a single strong peak, as observed
by Rotaciuc et al. (1991).

\section{Extinction}
\label{n1068:sec:ext}

The extinction towards individual H\,{\sc ii}~regions may not be uniform.
Young, Kleinmann \& Allen (1988) mapped the extinction using the ratio
of [S\,{\sc iii}]\,$\lambda$9532\AA \ to (H$\alpha$ + [N\,{\sc ii}]).
They found that it peaked at $A_{\rm V} = 1.5$\,mags to the
North-East, and at $A_{\rm V} = 2.5$\,mags to the South-West.
However, the assumption made in order to produce these maps has been
questioned by Kennicutt \& Pogge
(1990) who found an order of magnitude scatter in the intrinsic 
[S\,{\sc iii}]/H$\alpha$ ratio in H\,{\sc ii} regions.
They suggested that variations in the observed ratio might reflect
variations in the level of gas excitation rather than dust extinction, and
in the circumnuclear regions of NGC\,1068 several different line
excitation processes may be operating,
including photoionisation from the nucleus and
possibly relativistic electrons associated with the nuclear jet
(Cecil, Bland \& Tully 1990).

We have therefore measured the extinction in different regions by
comparing our Br$\gamma$ image to an H$\alpha$ image kindly
provided by J.~Bland-Hawthorn (Bland-Hawthorn et al. 1991).
This image shows {\it pure} H$\alpha$ emission, as
the contribution from [N\,{\sc ii}] has been mapped
and subtracted by fitting spectra at each pixel.
The image was calibrated by fixing the total line flux to be that
determined previously by these authors, $4.6 \times 10^{34}$\wsqm.
This was further verified by determing the difference in
(H$\alpha$+[N\,{\sc ii}]) flux (taking the average ratio
H$\alpha$/(H$\alpha$+[N\,{\sc ii}]) = 0.5) between a 13.5\,arcsec
aperture (McQuade, Calzetti \& Kinney, 1995) and a 64\,arcsec aperture
(Young et al. 1988).
These agreed to within 9~per cent.
The Br$\gamma$ image was smoothed to a similar resolution
(approximately 3\,arcsec FWHM) and re-binned to a scale of
0.5\,arcsec\,pixel$^{-1}$ to match the H$\alpha$ image.
The images were then registered with reference to several prominent
H\,{\sc ii} regions in the ring.
Br$\gamma$ and H$\alpha$ fluxes were measured in the regions A--O
using 3\,arcsec apertures.
The Br$\gamma$ fluxes were then velocity corrected as described in
Section~\ref{n1068:ss:vel} before deriving the extinction.
This was done assuming an intrinsic ratio H$\alpha$/Br$\gamma = 103.6$
(Osterbrock 1989) and using the Howarth (1983) galactic extinction
curve.
The results are given in Table~\ref{tab:n1068:prop}.

Typical values are $A_{\rm V} = 1$--3\,mag similar to those found
by Young et al. (1988), suggesting that reddening corrections to
infrared line fluxes are relatively small.
This result is important because by using a diagnostic in the K-band we
should be able to detect \hiiregions\ to optical depths much greater
than an \av\ of 3\,mag since $\tau_{2.2\mu \rm m} = 1$ at \av$ \simeq
9$\,mag
The issue cannot be avoided by considering, instead of the usual
`screen' model, a `mixed' model which allows one to infer
significantly higher \av\ but for which we still derive only \av$ =
3$--17\,mag.
It implies that in regions where we see star formation, either it is
occuring only near the surface, or the molecular clouds are not highly
optically thick.
The latter option seems unlikely from CO observations which imply
considerably higher optical thicknesses for the clouds, and the former
appears to require some external influence on the star formation.

\section{Timescales for Star Formation}
\label{n1068:sec:time}

\begin{table*}
\flushleft
\caption{Star Formation Properties \label{tab:n1068:prop}}
\begin{tabular}{ccccccccrrcc}

\\

Region & $A_{\rm V}$ & \multicolumn{2}{c}{Observed flux$^a$} &
\multicolumn{2}{c}{Corrected flux$^b$} & 2\um\ cont.$^c$
& S(1)/Br$\gamma$ & EW (Br$\gamma$) & $\log{N_{\rm UV}}$ 
& Age$^d$ & Mass$^d$\\

& mag & \multicolumn{2}{c}{$10^{-18}$\wsqm} &
\multicolumn{2}{c}{$10^{-18}$\wsqm} & mJy & ratio & \AA &
$10^{51}$\,s$^{-1}$ & Myr & 10$^6$\msun \\

&& Br$\gamma$ & 1-0\,S(1) & Br$\gamma$ & 1-0\,S(1) \\

\\

A & 1.2 & $3.3\pm0.2$ & $3.5\pm0.3$ & $3.7\pm0.2$ & $3.9\pm0.3$ & 
$3.24\pm0.03$ & $1.0\pm0.1$ & $ 16\pm1.0$ &  6.6 & 6.5 & 1.1 \\

B & 2.0 & $4.3\pm0.2$ & $3.2\pm0.3$ & $5.1\pm0.3$ & $3.9\pm0.4$ & 
$2.27\pm0.03$ & $0.8\pm0.1$ & $ 29\pm1.5$ &  9.1 & 6.1 & 1.1 \\

C & 1.9 & $2.7\pm0.2$ & $1.8\pm0.3$ & $3.2\pm0.3$ & $2.2\pm0.4$ & 
$1.98\pm0.03$ & $0.7\pm0.1$ & $ 21\pm1.9$ &  5.7 & 6.2 & 0.8 \\

D & 2.0 & $3.6\pm0.2$ & $2.3\pm0.3$ & $4.3\pm0.3$ & $2.7\pm0.4$ & 
$1.24\pm0.03$ & $0.6\pm0.1$ & $ 45\pm3.3$ &  7.6 & 5.8 & 0.8 \\

E & 3.2 & $3.4\pm0.4$ & $1.7\pm0.5$ & $4.6\pm0.5$ & $2.3\pm0.7$ &
$0.65\pm0.03$ & $0.5\pm0.2$ & $ 83\pm11 $ &  8.2 & 5.3 & 0.6 \\

F & 1.1 & $5.7\pm0.5$ & $< 2.2$     & $6.4\pm0.6$ & $< 2.4$ & 
$1.91\pm0.03$ & $<0.4$      & $ 47\pm4.5$ & 11.2 & 5.8 & 1.2 \\

G & 1.3 & $3.6\pm0.5$ & $< 2.1$     & $4.1\pm0.6$ & $< 2.3$ & 
$3.84\pm0.03$ & $<0.6$      & $ 15\pm2.1$ &  7.2 & 6.6 & 1.2 \\

H & 1.5 & $3.1\pm0.4$ & $< 1.4$     & $3.6\pm0.4$ & $< 1.7$ & 
$1.09\pm0.03$ & $<0.5$      & $ 45\pm5.2$ &  6.6 & 5.8 & 0.7 \\

I & 2.8 & $2.8\pm0.3$ & $1.6\pm0.4$ & $3.7\pm0.4$ & $2.1\pm0.6$ & 
$0.44\pm0.03$ & $0.6\pm0.2$ & $101\pm13 $ &  6.7 & 4.4 & 0.4 \\

J & 2.0 & $1.8\pm0.3$ & $2.0\pm0.4$ & $2.2\pm0.3$ & $2.4\pm0.5$ & 
$0.10\pm0.03$ & $1.1\pm0.3$ & $282\pm100$ &  3.9 & 4.3 & 0.2 \\

K & 2.9 & $1.5\pm0.3$ & $< 1.0$     & $1.9\pm0.3$ & $< 1.4$ & 
$0.12\pm0.03$ & $<0.8$      & $185\pm58 $ &  3.4 & 4.3 & 0.2 \\

L & 3.0 & $1.9\pm0.2$ & $1.6\pm0.3$ & $2.5\pm0.3$ & $2.1\pm0.4$ & 
$0.19\pm0.03$ & $0.8\pm0.2$ & $154\pm31 $ &  4.5 & 4.3 & 0.3 \\

M & 1.7 & $2.0\pm0.2$ & $< 1.0$     & $2.3\pm0.3$ & $< 1.1$ & 
$0.27\pm0.03$ & $<0.5$      & $117\pm20 $ &  4.1 & 4.4 & 0.2 \\

N & 2.1 & $5.1\pm0.2$ & $1.8\pm0.3$ & $6.2\pm0.3$ & $2.2\pm0.4$ & 
$0.45\pm0.03$ & $0.4\pm0.1$ & $176\pm15 $ & 10.8 & 4.3 & 0.6 \\

O & 0.5 & $5.0\pm0.2$ & $1.7\pm0.3$ & $5.2\pm0.2$ & $1.8\pm0.3$ & 
$0.55\pm0.03$ & $0.3\pm0.1$ & $143\pm10 $ &  9.3 & 4.3 & 0.5 \\

\\

\end{tabular}

Quoted errors are 1$\sigma$ and limits are 3$\sigma$.\\
$^a$ Measured in a 3\,arcsec aperture and corrected for the velocity
field (see Section~\ref{n1068:ss:vel}).\\
$^b$ Corrected for the extinction listed in column~2 using 
$A_{\rm K}/A_{\rm V} = 0.1$.\\
$^c$ Observed flux density, measured in a 3\,arcsec aperture.\\
$^d$ Assuming instantaneous star formation.

\end{table*}

The spatial resolution in Fig.~\ref{fig:n1068:morph} has allowed us to
measure the FWHM of some Br$\gamma$ knots as 1.3\,arcsec (90\,pc), and
in the unsmoothed image we have measured the brightest knot to have a
FWHM of 0.9\,arcsec (60\,pc).
These upper limits to the cluster sizes, and the masses we derive in
this section, suggest that each region A--O,
if not an individual OB association, is probably a relatively small
number of such associations.

We can determine the age of the star formation by measuring the
equivalent width (EW) of the Br$\gamma$ line in these clusters.
In order to estimate the continuum flux density associated only with
young stars we have subtracted the underlying isophotes
(Fig.~\ref{fig:n1068:morph}).
Careful comparison of the images suggests the labelled regions even close to
the bar will suffer little or no contamination from the continuum in
the bar.
The Br$\gamma$ and continuum fluxes in 3\,arcsec apertures at these
sites are given in Table~\ref{tab:n1068:prop}, and show that
EW(Br$\gamma$)=20--200\AA.
We use the evolutionary synthesis models of Leitherer \& Heckman (1995)
to determine the starburst history, making the assumptions of solar
metallicity and a standard IMF (Salpeter in the range 1--100\msun).

If the star formation is instantaneous (i.e., the duration is small
compared to the time elapsed since it occured) then
around the whole ring, it occured from 4--7\,Myr ago.
The similarity in the ages even with an order of magnitude
difference in EW is due to the very rapid decrease in EW with time,
reflecting the speed of evolution of the most massive
stars away from the main sequence.
The ionising flux $N_{\rm UV}$ is derived from the dereddened Br$\gamma$
flux making the usual assumption that 70 Lyman continuum photons are
required for each Br$\gamma$ photon produced.
This and the age can then be used to determine the mass of stars formed
in each region, which is 10$^5$--10$^6$\msun.

Alternatively, if the star formation is still progressing, then the
ages range from 6\,Myr to more than 300\,Myr (unless the IMF is
truncated).
Although the total mass of stars formed would be of the order of
10$^6$--10$^7$\msun\ in each complex, this is still several times
smaller than the available gas mass (estimated from the CO luminosity,
Planesas et al. 1991) and so does not rule out the scenario.
We believe that continuous star formation is unlikely as it would need to
be sustained over timescales of more than 100\,Myr in highly specific
regions.
In a single star cluster, once massive young stars have formed,
the turbulence and heating from these would begin to disrupt the
remainder of the giant molecular cloud (GMC) and inhibit further star
formation.
After a few million years these stars would evolve to produce
supernovae, injecting more kinetic energy into the local ISM.
Further star formation would be halted after only a few generations of
OB stars.
Blitz \& Shu (1980) and Blitz (1991) used observational and theoretical
evidence to argue that GMCs which generate OB
associations cannot have lifetimes much greater than 20--40\,Myr.
A simple persuasive line of reasoning concerns the gravitational
binding energy of a GMC, typically $10^{50}$\,erg for a $2 \times
10^5$\msun\ cloud of size 40\,pc.
By estimating the luminosity of OB stars formed over $10^7$\,yrs and
the efficiency with which this could be converted into kinetic energy
Blitz \& Shu found that $10^{51}$\,erg were available over this time,
which could produce expansion speeds of $\sim 10$\kms\ capable of
dispersing the cloud.
Thus even without considering the kinetic energy released by supernovae
(typically $10^{51}$\,erg for a type~II supernova), the active star
formation episode in GMCs is necessarily short lived.
More detailed calculations of the mechanical energy deposited in the
ISM are included in Leitherer \& Heckman's models and we apply these to
NGC\,1068.
They show that for a cluster undergoing constant star
formation with the typical Br$\gamma$ luminosity observed, it
exceeds the binding energy of $\sim10^{55}$\,erg  for the larger
molecular clouds (Planesas et al. 1991) after 30--40\,Myr.
This therefore provides an upper limit to the duration of the star
forming episodes here.
The masses of the clouds may actually be much less as there is some
evidence that the CO-to-H$_2$ conversion factor has been overestimated,
in which case the the binding energy would be
significantly smaller and the limit on the duration of star formation
could be less than 10--20\,Myr. 

We conclude that the active phase in the ring is
short-lived, these are young star clusters, and that the star formation
is co-eval on a timescale of 10\,Myr.
The 2\um\ continuum in these regions can be accounted for entirely by
this star formation, so that we find no evidence for previous episodes
of star formation.
Since strong H$\alpha$ is observed around the whole ring, the arguments
above can be extended to the regions where we could not detect
Br$\gamma$.
Even to the West where we observe continuum emission
(0.1\,mJy\,arcsec$^{-2}$) but no Br$\gamma$ ($< 0.22 \times
10^{-18}$\wsqma\ averaged over a 3\,arcsec aperture and velocity
corrected), the upper limit for EW(Br$\gamma$) is 35\AA\ consistent
with an age of 6\,Myr.
We therefore also conclude that this is probably the first episode of star
formation in the ring.

\section{Triggering Star Formation}
\label{n1068:sec:sf}

The existence of the circumnuclear ring in NGC\,1068 has been
attributed to gas settling between the inner Lindblad resonances (ILRs)
as a result of the action of a barred gravitational potential (Telesco
\& Decher 1988).
In such situations we expect vigorous star formation to occur as a direct
result of the increased cloud density at the ILR. 
The Schmidt Law describes the star formation rate as dependent
on the local gas surface density ($SFR \propto \rho_{\rm gas}^n$
with $1 < n < 2$, eg Kennicutt 1989)
because the higher cloud density leads to an increased cloud
collision rate so that GMCs are more likely to form
and compression induced cloud collapse to occur, both of which will
enhance the star formation rate.
This appears to be substantiated, at least qualitatively, by the
alignment of H$\alpha$, Br$\gamma$, 10\um, 2\um, and CO emission at the
ILR.

The radio jet to the North-East extends only 6\,arcsec from the nucleus
and so has not yet impinged on the ring (Wilson \& Ulvestad 1983), but
it is interesting to speculate on what effect it may have if this happens.
Models such as those proposed by Rees (1989) suggest that once a jet's
bow shock has passed, clouds will be subjected to an overpressure of
$\sim {\cal M}^2$ where $\cal M$ is the Mach number.
From the half-apex angle of the cone which they interpreted as the bow
wave generated by the jet's motion, Wilson \& Ulvestad (1983) estimated
that ${\cal M} \simgt 3$, sufficient to trigger further star formation.

The ionisation cone from the AGN may have an effect on the
environment to the North-East of the nucleus since it extends up to and
beyond the ring, and may be assisting the star formation by heating
the ISM so that the overpressure helps to crush the clouds.
Similar mechanisms have been put forward for jet-induced star formation
where the hot plasma behind the shock front overpressures the cloud
triggering star formation (eg van Breugel et al. 1985; Best, Longair \&
R\"{o}ttgering 1997).
We consider whether this mechanism may apply here.
When clouds first drift in from the galactic disc, they will probably
have temperatures $T \sim 10$\,K and average
densities $n_{\rm e} \sim 10^3$\pcmcu;
after some time in the environment of the inner galaxy, they
will heat up and perhaps coalesce to form bigger, denser clouds
with $T \sim 100$\,K and $n_{\rm e} \sim 10^4$\pcmcu\ (Goldsmith 1987;
Genzel 1991).
These clouds will have pressures $P/k$ of $10^4$\,K\pcmcu\ and
$10^6$\,K\pcmcu\ respectively.
Once star formation has commenced H\,{\sc ii} regions can provide the
overpressure to propagate star formation throughout the remaining GMCs.
At least some of the gas in the ionisation cone photoionised by the
Seyfert nucleus is in a high excitation state.
This has been modelled by a number of authors (Evans \& Dopita 1986;
Bergeron et al. 1989; Nazarova 1994) to have  
$T = 3$--$5 \times 10^4$\,K and $n_{\rm e} = 0.1$--1\pcmcu.
The pressure due to this phase is then $P/k \sim 10^4$\,K\pcmcu,
clearly not high enough to affect the GMCs in the current epoch.
On the other hand, the soft X-ray morphology, which is aligned toward
the North-East, has been attributed to a nucleus-driven wind with a
pressure of 2--$8 \times 10^6$\,K\pcmcu\ (Wilson et al. 1992).
These authors suggest this could confine the narrow-line clouds, and 
would also be sufficient to provide some overpressure on clouds in
parts of the ring, perhaps providing an external stimulous for star
formation as required in Section~\ref{n1068:sec:ext}

\section{The Role of the Nucleus}
\label{n1068:sec:nucleus}

\begin{figure}
\plotfiddle{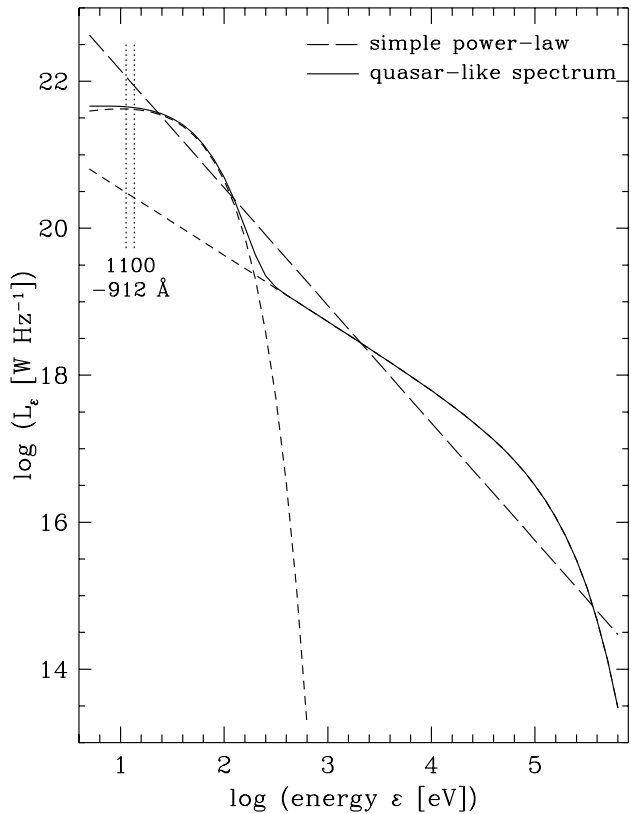}{11cm}{0}{43}{43}{-125}{2}
\caption{Spectra of the two models of the nuclear luminosity, scaled
to the estimated ionising flux of $Q = 8 \times 10^{54}$\,sec$^{-1}$.
That used by Storchi-Bergmann et al. (1992) is a simple power-law,
while that from Bland-Hawthorn et al. (1997) has a more realistic
quasar form, incorporating a power-law with a turnover as well as a
`big blue bump' component.}
\label{n1068:fig:spectrum}
\end{figure}

Bland-Hawthorn et al. (1997) have proposed a model in which the nuclear
continuum luminosity incorporates a big blue bump with a turnover at
$\varepsilon_1 = 30$\,eV
\[
L_\varepsilon ({\rm photons\,s^{-1}\,eV^{-1}}) = 
k_1 \varepsilon^{-2/3} \exp{[-\varepsilon/\varepsilon_1]}
\]
as well as a power-law with slope $\alpha = 1.9$ turning over at
$\varepsilon_1 = 100$\,keV 
\[
L_\varepsilon ({\rm photons\,s^{-1}\,eV^{-1}}) = 
k_2 \varepsilon^{-\alpha} \exp{[-\varepsilon/\varepsilon_2]}
\]
($k_1$ and $k_2$ are constants) 
which add together to make a more realistic quasar spectrum of the form
shown in Fig.~\ref{n1068:fig:spectrum}.
This luminosity heats grains at the
circumnuclear ring which are lying within the ionisation cones.
The small grains which can easily be heated to high temperatures would
be responsible for the enhanced mid-infrared continuum while the larger
grains would re-radiate at longer wavelengths.
Their model predicts continuum fluxes over a range of wavelengths from
2--100\um, and was envisaged in order to explain the 10\um\ emission
in the ring near the ends of the bar (Telesco \& Decher 1988).

In the previous sections we considered only the role of star formation
for the emission processes in the ring.
But the continuum-residual image 
(Fig~\ref{fig:n1068:morph}) shows that the strongest 2\um\ emission not
associated with the bar coincides almost exactly with the 10\um\
emission, and hence may be excited by the same mechanism.
We are unable to impose any useful constraints on such a model as the
predicted total 2\um\ flux is only 1\,mJy, considerably less than that
observed at the same positions.

\subsection{Circumnuclear H$_2$ Emission}
\label{n1068:sec:h2}

\begin{figure}
\psfig{file=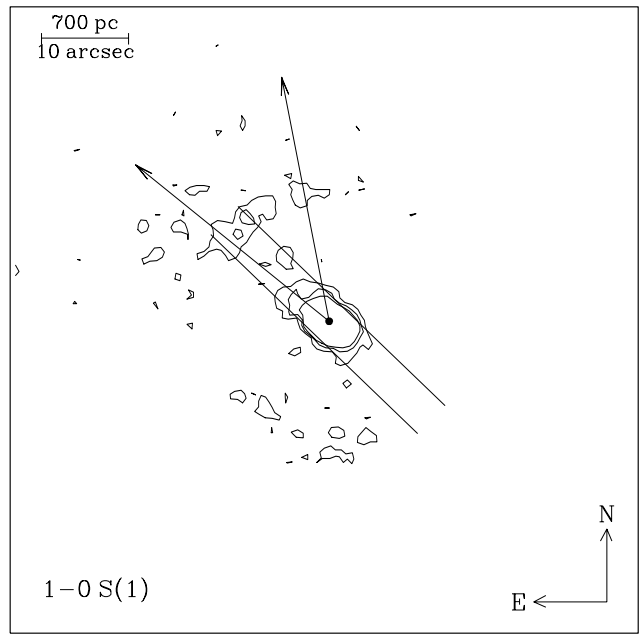,width=8.4cm,rheight=8.4cm}
\caption{Contour plot of 1-0\,S(1) in NGC\,1068.
Contour intervals are as for Fig.~1, and are not corrected for the velocity
field.
The parallel lines mark the extent of the bar at at 0.5\,mJy and
position angle 46\deg.
The arrows delineate the extent of the ionisation cone as described by
Unger et al. (1992), at PA 11\deg\ and PA 51\deg.
The cone extends beyond the ring.}
\label{fig:n1068:cone}
\end{figure}

Fig.~\ref{fig:n1068:cone} shows the alignment of the bar (parallel
lines) and the ionisation cone (diverging arrows) with respect to the
observed 1-0\,S(1) emission, although direct interpretation of the figure
should be approached cautiously due to the calibration difficulties.
The dereddened fluxes in
Table~\ref{tab:n1068:prop} show that the intensity of the
Br$\gamma$ flux is similar at both ends of the bar.
On the other hand, there is a significant difference in the 1-0\,S(1)
fluxes:
that at the North end of the bar is almost twice that at the South end.
This is reflected in the S(1)/Br$\gamma$ ratio which is 0.7--1.0 to the
North and $\simlt 0.5$ to the South.
Note that the high S(1)/Br$\gamma$ ratios observed in two other
regions (J and L) are due to weak Br$\gamma$ rather than
strong 1-0\,S(1);
In every region except A--D the 1-0\,S(1) flux is less than $2.5 \times
10^{-17}$\wsqm, while in regions A and B it reaches $3.9 \times
10^{-17}$\wsqm.
We propose that there may be an additional excitation source for the H$_2$
in these regions.

Puxley, Hawarden \& Mountain (1990) proposed the use of the
S(1)/Br$\gamma$ ratio as a diagnostic for models of star forming
regions, assuming that the 1-0\,S(1) was excited by fluorescence.
Their model~B, a compact stellar cluster surrounded by a single large
H\,{\sc ii} region, predicts a ratio of 0.1--0.4 for electron densities
$\simgt 100$\pcmcu\ and photodissociation region gas densities
10$^3$--10$^4$\pcmcu\ and $N_{\rm UV} = 10^{51}$--10$^{52}$\,s$^{-1}$.
Their model~A, involving less compact clusters in which each star has
its own H\,{\sc ii} region, produces higher ratios for the same gas
densities.
However, there is considerable evidence that stars form in compact
clusters (eg Meurer et al. 1995), which would tend to result in a low
S(1)/Br$\gamma$ ratio for purely fluorescent H$_2$ emission, so that
for the ratios observed here fluorescence is unlikely to be the only
source of 1-0\,S(1) flux.

Supernova remnants, appearing in a starburst after 3\,Myr,
will produce some emission by shocking the ISM;
also, where there is such a high density of clouds, shock excitation
directly via cloud collisions may also be an important contributor.
Together these may explain the observed tendency towards 
S(1)/Br$\gamma \sim 0.5$ .
However, all the processes mentioned so far will follow either the star
cluster or cloud distributions which are fairly uniform around the ring.
To account for the large difference in 1-0\,S(1) flux at opposite ends
of the bar, we propose that the extra emission at the North
end, 1--$2 \times 10^{-18}$\wsqm\ in the 3\,arcsec apertures, is due
to a process unrelated to star formation and cloud collisions.
One possibility is fluorescence at the edges of the molecular clouds by
UV radiation from the nucleus.
That the nucleus has an important influence on this region is
unequivocally demonstrated by the [O\,{\sc iii}] ionisation cone (Unger
et al. 1992).
Bergeron et al. (1989) showed that the emission line spectrum in this
region could be modelled in terms of photoionisation by soft X-rays
from the nucleus.
We argue that the UV radiation from the nucleus may also be affecting
the line excitation within the molecular clouds in this direction.

In order to quantify this effect, we convert the incident
UV flux into an emitted 1-0\,S(1) intensity, using the detailed models
of H$_2$ excitation from Black \& van~Dishoeck (1987).
These require us to estimate the 912--1100\,\AA\ flux, which we can do
using the same prescription that Storchi-Bergmann, Mulchaey, \& Wilson
(1992) used to calculate the luminosity of the torus in NGC\,1068.
The starting point is the observed characteristics of the
ionisation cone, which has a direct line of sight to the nucleus.
Then by representing the nuclear continuum emission as a power-law $L_\nu =
A \nu^{-\alpha}$, we can derive the coefficient $A$, and estimate the
912--1100\,\AA\ flux incident at the circumnuclear ring
(Fig.~\ref{n1068:fig:spectrum}).
Storchi-Bergmann et al. calculate the number of ionising photons $Q$
from the ionisation parameter $U$ 
(from [O\,{\sc iii}]\,$\lambda$5007/H$\beta$ and 
[N\,{\sc ii}]\,$\lambda$6583/H$\alpha$) and gas density $n$ 
(from the ratio of the [S\,{\sc ii}] lines) measured at
several places in the ionisation cone at distances $r$ from the nucleus:
\[
Q = U\,4 \pi r^2 n c
\]
The error estimates lead to an uncertainty in nuclear luminosity of a
factor of 8, but typical values are $U = 10^{-3}$ and $n =
150$\cmcu\ at $r = 4$\,kpc.
The coefficient $A$ is found by integrating over all $\nu \ge \nu_0$,
where $\nu_0$ is the frequency corresponding to the Lyman limit of
912\,\AA, giving $A  = Q h \alpha \nu_0^\alpha$.
The spectral index was measured by Kinney at al. (1991) to be $\alpha =
1.6$, similar to the typical Seyfert~1 value of 1.5.
The 912--1100\,\AA\ luminosity is simply
\[
L_{\rm UV} = \int^{\nu_2}_{\nu_1} A \nu^{-\alpha} d\nu
= Q h \alpha \nu_0^\alpha (\nu_2^{1-\alpha} - \nu_1^{1-\alpha}) / (1 - \alpha)
\]
where $\nu_1$ and $\nu_2$ are the frequencies for 1100 and 912\,\AA\ 
respectively.
Similarly, integrating over all $\nu \ge \nu_0$ gives the total
ionising luminosity as $5 \times 10^{37}$\,W.

Sokolowski, Bland-Hawthorn \& Cecil (1991) use a similar method but
proceed in the opposite direction:
they begin by assuming a luminosity for the nucleus and model the
interstellar medium by fitting the observed line profiles.
They estimate $10^{-3.1} \le U \le 10^{-2.6}$ on the ionisation cone and
$10^{-4.5} \le U \le 10^{-4.1}$ away from it for $r = 2$--4\,kpc,
finding it optimal to set $n = 30$\pcmcu\ at 2\,kpc, decreasing
as $r^{-1}$.
However, their estimate of the nuclear luminosity from X-ray
observations does not take account of internal extinction and ignores
the possibility of significant flux from photons with E$\simgt 15$\,keV.
As such it is about 1/4 that of both the nuclear infrared luminosity
(Telesco et al. 1984) and the estimated nuclear bolometric luminosity
(Pier et al. 1994).
If we corrected for this, the numerical factor would carry through
the model and increase their parameter $U \times n$ by a
similar amount, making it consistent with that of
Storchi-Bergmann et al.

Using the values above, we find $L_{\rm UV} = 6 \times 10^{36}$\,W.
Since there must be rather little extinction in the
direction of the ionisation cone (we assume none) the UV flux
incident at the circumnuclear ring, 15.5\,arcsec from the
nucleus, is $4.4 \times 10^{-4}$\wsqm, nearly 3 orders of magnitude
higher than the mean background in the solar neighbourhood of 
$6 \times 10^{-7}$\wsqm (Black \& van Dishoeck 1987).
Lastly, if the molecular clouds have densities of $\sim 10^4$\pcmcu\
and we take the covering factor of the clouds to be unity (Planesas et
al. 1991 find it to be significantly higher than for Galactic GMCs)
then in a 3\,arcsec region the models of Black \& van Dishoeck give the
1-0\,S(1) flux as $4 \times 10^{-18}$\wsqm.

Two important points to consider are: 
(1) the factor 8 uncertainty from Storchi-Bergamnn et al makes little
difference in practice as in all cases the UV intensity would still be
more than 100 times the background intensity, and the resulting
1-0\,S(1) line flux would be reduced by at most a factor 2;
(2) Sokolowski et al.'s model indicates that the UV intensity away
from the ionisation cones is 30 times less than in them, and is
insufficient to cause fluorescent H$_2$ emission.

We also consider whether this scenario is compatible with the
quasar-nucleus model of Bland-Hawthorn et al. (1997) described at the
beginning of the section.
The quasar spectrum they employed, scaled to the same $Q$ that we
calculated previously as shown in Fig.~\ref{n1068:fig:spectrum},
results in an ionising luminosity the same as for the simple power law.
But the 1100--912\AA\ luminosity is a factor of 2 less (as the spectrum
is essentially flat at these energies), although within the
uncertainty of our previous derivation and sufficient to excite
1-0\,S(1) fluorescence at the ring.
In order to explain the 10\um\ emission, Bland-Hawthorn et
al. required an EUV (10\,eV -- 10\,keV)
luminosity of $3 \times 10^{38}$\,W, a factor of 6 more than the
ionisation and density parameters suggest.
However, the resulting 1-0\,S(1) emission is not significantly
enhanced.

We have demonstrated that in principle a nuclear continuum
which reproduces the characteristics of the ionisation cone can also
cause fluorescence at the circumnuclear ring, and that this mechanism
accounts for both the extra flux observed as well as the restricted
location.

\section{Near-Nuclear H$_2$ Emission}
\label{n1068:sec:inner}

We include a short section here on the nuclear H$_2$ emission as our
data show, due to considerably lower detection limits and wider field,
that the emission is more extended than was previously realised.
Hall et al. (1981) found that the emission line spectrum indicated
the H$_2$ was excited thermally with $T_{\rm ex} \simlt 2500$\,K
concluding, as did Oliva \& Moorwood (1990), that it should be
associated with stellar processes.
On the other hand, Rotaciuc et al. (1991) and Maloney (1997) argued
that a nonstellar source of ionising radiation such as an X-ray
emitting AGN, could be responsible for heating the clouds.
Blietz et al. (1994) suggested that there was a close physical
connection between the H$_2$ clouds and radio jet, although the nuclear
H$_2$ knot could not represent the torus proposed by Antonucci \& Miller
(1985).

Our observations are not of a sufficiently high resolution to shed new
light on this debate.
However, they do show that weak H$_2$ emission extends for 10\,arcsec
(700\,pc) along the bar.
This is more remarkable because the intensity falls away extremely
quickly perpendicular to this axis, more than an order of magnitude in
1\,arcsec.
Such a morphology is unlikely to arise as a result of collimated
emission from the nucleus, such as
fluorescence due to X-ray irradiation of molecular clouds.
This mechanism would only be viable in the directions along
which the high energy photons can escape, that is within the
ionisation cone.
We also consider that star forming processes cannot be responsible as
the Br$\gamma$ flux has a rather different morphology and the
S(1)/Br$\gamma$ ratio is $\simgt 1$.
Both of these are contrary to expectations of star forming regions.

The alternative we propose is that, since the emitting region appears to
coincide with the bar, the excitation mechanism is probably connected
with it.
The bar inside the circumnuclear ring is a decoupled secondary and
typically the co-rotation radius of an inner bar coincides with the
ILRs of the primary bar.
Helfer (1997) measured the inner rotation curve and found that if
corotation occurs at $r=15$\,arcsec, the pattern speed is 
160\kms\,kpc$^{-1}$, and that there may be inner Lindbald resonances
associated with it in the inner hundred parsecs.
Consequently, for gas at radii $r < 15$\,arcsec inflow can proceed
right into the nuclear region.
Athanassoula (1992) discusses the response of gas to barred potentials,
showing that shocks occur along the leading edge of a bar where the gas
flow is close to radial.
These models do qualitatively describe the inner regions of NGC\,1068
in which the gas is marginally offset from the bar (to the East of the
North-East arm and West of the South-West arm, Helfer 1997), and as
such may help in understanding the observed H$_2$ at radii 
$r < 15$\,arcsec which shows weak 1-0\,S(1) emission along part of the
bar, with a morphology entirely consistent with the CO map.
Even though kinematic studies of the inner disc have shown that there
may be a general expansion at velocities 50--100\kms\ (eg Atherton et
al. 1985, Planesas et al. 1991), we do not consider that this would
compromise a scenario in which infall is occuring along the bar:
clouds form a highly dissipative medium and the loss of angular
momentum suffered by clouds in the shock would be sufficient to
overcome any reasonable outward accelerating effect, with the relatively
high density of clouds in the bar resulting in a net infall.
Spatially and velocity resolved observations of HCN (Tacconi et
al. 1994) provide further evidence that the bar influences the
kinematics of the inner disc, and that gas is streaming along highly
elongated orbits. 
These authors concluded that the H$_2$ in this region was shock excited,
and that dissipation was leading to influx to the nucleus at the rate
of a few \msun\,yr$^{-1}$ and a speed of $\sim$50\kms.

\section{Conclusions}
\label{n1068:sec:conc}

We have presented the first near-infrared line images of NGC\,1068 to
include the circumnuclear ring.
The fluxes are calibrated to take into account the strong variations in
the velocity field across the inner few kpc, which meant
that the line flux transmitted by the Fabry-Perot also varied across
the image.

We have derived the extinctions in a number of star forming knots by
measuring the H$\alpha$/Br$\gamma$ recombination line ratios.
These are typically in the range $A_{\rm V} = 1$--3\,mag.
Comparison of the absolute line fluxes and the continuum due to young
stars (obtained by subtracting the underlying elliptical component) to
evolutionary synthesis models
shows that the star formation in the ring occured during a short burst
within the last 30--40\,Myr, and perhaps as recently as 4--7\,Myr ago;
and that this was the first such episode of star
formation in the ring.
Although the ionisation cone crosses the North-East segment of the
ring, it cannot have a substantial effect on the star formation.
On the other hand, the X-ray wind may do so, and once the radio jet
reaches the ring it is likely that this will also affect it.

The sizes and masses of the Br$\gamma$ knots indicate that the stars
are forming in compact clusters.
The 1-0\,S(1) fluxes and S(1)/Br$\gamma$ ratios in these regions
suggest some component of the 1-0\,S(1) emission is due to
shock excitation from cloud collisions or supernova remnants.
However, there is an excess flux to the North-East, coincident with
the region in which the ionisation cone crosses the ring.
We have investigated whether this could be due to fluorescence at the
edges of molecular clouds by UV radiation from the Seyfert nucleus which
is also producing the high excitation lines, and conclude that such a
process is possible.

We have found that the nuclear H$_2$ emission is more extended than was
previously realised, and that this occurs only along the major axis of
the bar.
It is possible that this is associated with shocks induced by the
streaming motions of clouds along the leading edge of the bar.

\section*{Acknowledgments}

We would like to thank staff members at UKIRT for their help in
operating the telescope, and Joss Bland-Hawthorn for kindly providing
us with his H$\alpha$ image, and as referee for his useful comments and
suggestions.
This research was supported by a PPARC (EPSRC) research studentship grant.


\bsp

\label{lastpage}


\begin{thebibliography}{}

\bibitem{am85}
Antonucci R., Miller J., 1985, ApJ, 297, 621

\bibitem{art85}
Atherton P., Reay N., Taylor K., 1985, MNRAS, 216, 17P

\bibitem{a92}
Athanassoula E., 1992, MNRAS, 259, 345

\bibitem{bpd89}
Bergeron J., Petitjean P., Durret F., 1989, A\&A, 213, 61

\bibitem{blr97}
Best P., Longair M., {R\"{o}ttgering} H., 1997, MNRAS, 286, 785

\bibitem{bd87}
Black J., van~Dishoeck E., 1987, ApJ, 322, 412

\bibitem{bsc91}
Bland-Hawthorn J., Sokolowski J., Cecil G., 1991, ApJ, 375, 78


\bibitem{bgt97b}
Bland-Hawthorn J., Voit G., Cecil G., Weisheit J., 1997, Ap\&SS, 248,
177

\bibitem{bcd94}
Blietz M., Cameron M., Drapatz S., Genzel R., Krabbe A., van der Werf
P., Sternberg A., Ward M., 1994, ApJ, 421, 92

\bibitem{b91}
Blitz L., 1991, in The Physics of Star Formation and Early Stellar
Evolution, eds Lada C., Kylafis N., ASI Series vol.342, p.3, Kluwer
Academic Publishers, Dordrecht

\bibitem{bs80}
Blitz L., Shu F., 1980, ApJ, 238, 148


\bibitem{cbt90}
Cecil G., Bland J., Tully R., 1990, ApJ, 355, 70

\bibitem{ed86}
Evans I., Doptia M., 1986, ApJ, 310, L15


\bibitem{gbo07}
Gallimore J., Baum S., O'Dea C., 1997, Ap\&SS, 248, 253

\bibitem{g91}
Genzel R., 1991, in The Physics of Star Formation and Early Stellar
Evolution, eds Lada C., Kylafis N., ASI Series vol.342, p.155, Kluwer
Academic Publishers, Dordrecht

\bibitem{g87}
Goldsmith P., 1987, in Interstellar Processes, eds Hollenbach D.,
Thronson H., Reidel (Dordrecht), p.51

\bibitem{hks81}
Hall D., Kleinmann S., Scoville N., Ridgway S., 1981, ApJ, 248, 898

\bibitem{h97}
Helfer T., 1997, Ap\&SS, 248, 51

\bibitem{h83}
Howarth I., 1983, MNRAS, 203, 301

\bibitem{kw78}
Keel W., Weedman D., 1978, ApJ, 192, 581

\bibitem{k89}
Kennicutt R., 1989, ApJ, 344, 685

\bibitem{kp90}
Kennicutt R., Pogge R., 1990, AJ, 99, 61

\bibitem{kaw91}
Kinney A., Antonucci R., Ward M., Wilson A., Whittle M., 1991, ApJ,
377, 100

\bibitem{lh95}
Leitherer C., Heckman T., 1995, ApJS, 96, 9

\bibitem{m97}
Maloney, P., 1997, Ap\&SS, 248, 105


\bibitem{mck95}
McQuade K., Calzetti D., Kinney A., 1995, ApJS, 97, 331

\bibitem{n94}
Nazarova L., 1994, in Proc. of the Oxford Torus Workshop, ed Ward M.,
University of Oxford, p.31

\bibitem{om90}
Oliva E., Moorwood A., 1990, ApJ, 348, L5

\bibitem{o89}
Osterbrock D., 1989, in Astrophysics of Gaseous Nebulae and Active
Galactic Nuclei. University Science Books, Mill Valley, USA.

\bibitem{pah94}
Pier E., Antonucci R., Hurt T., Kriss G., Krolik J., 1994, ApJ, 428, 124

\bibitem{psm91}
Planesas P., Scoville N., Myers S., 1991, ApJ, 369, 364

\bibitem{p88}
Pogge R., 1988, ApJ, 328, 519

\bibitem{phm90}
Puxley P., Hawarden T., Mountain C., 1990, ApJ, 364, 77

\bibitem{raw86}
Reay N., Atherton P., Walton N., 1986, MNRAS, 218, 13P

\bibitem{r89}
Rees M., 1989, MNRAS, 239, 1P

\bibitem{rkc91}
Rotaciuc V., Krabbe A., Cameron M., Drapatz S., Genzel R., Sternberg
A., Storey J., 1991, ApJ, 370, L23

\bibitem{smc88}
Scoville N., Matthews K., Carico D., Sanders D., 1988, ApJ, 327, L61


\bibitem{smw92}
Storchi-Bergmann T., Mulchaey J., Wilson A., 1992, ApJ, 395, L73

\bibitem{sbc91}
Sokolowski J., Bland-Hawthorn J., Cecil C., 1991, ApJ, 375, 583

\bibitem{tgb94}
Tacconi L., Genzel R., Blietz M., Cameron M., Harris A., Madden S.,
1994, ApJ, 426, L77

\bibitem{tbw84}
Telesco C., Becklin E., Wynn-Williams C., Harper D., 1984, ApJ, 282, 427

\bibitem{tqg97}
Thatte N., Quirrenbach A., Genzel R., Maiolino R., Tecza M., 1997, ApJ,
490, 238

\bibitem{td88}
Telesco C., Decher R., 1988, ApJ, 334, 573

\bibitem{ulp92}
Unger S., Lewis J., Pedlar A., Axon D., 1992, MNRAS, 258, 371

\bibitem{bfh85}
van~Breugel W., Filippenko A., Heckman T., Miley G., 1985, ApJ, 293, 83

\bibitem{wu83}
Wilson A., Ulvestad J., 1983, ApJ, 275, 8

\bibitem{wel92}
Wilson A., Elvis M., Lawrence A., Bland-Hawthorn J., 1992, ApJ, 391, L75

\bibitem{yka88}
Young  J., Kleinmann S., Allen L., 1988, ApJ, 334, L63


\end{thebibliography}
\end{document}